\documentclass
[prd,twocolumn,showpacs,preprintnumbers]{revtex4}
\usepackage{amssymb}
\usepackage{dcolumn}% Align table columns on decimal point
\usepackage{bm}% bold math
\usepackage{amsfonts}
\usepackage{latexsym}
\usepackage{amsmath}

\usepackage[dvips]{epsfig}
\newcommand{\nn}{\nonumber}
\newcommand{\nnn}{\nonumber \\ }
\newcommand{\lal}{&}
\begin{document}

\title{
Radiation reaction in curved even-dimensional spacetime}
\author{Dmitri V. Gal'tsov, Pavel Spirin \footnote{\tt E-mail: galtsov@physics.msu.ru,
salotop@list.ru}} \affiliation{Department of Theoretical Physics,
Moscow State University, 119899, Moscow, Russia}

%\maketitle
\begin{abstract}
We develop a new method of computing radiation reaction for a point
particle interacting with massless scalar and vector fields in
curved space-time. It is based on the analysis of field Green's
functions with both points lying on the particle world-line and does
not require integration of the field stresses outside the world line
as was used in the DeWitt-Brehme approach, thus leading to a
substantial simplification of the problem. We start with space-time
of an arbitrary dimension  and show that the Hadamard expansion of
the massless scalar and vector Green's functions contain only
integer inverse powers of the Synge world function in even
dimensions and only half-integer in the odd dimensions. The
even-dimensional case then is treated in  detail. We analyze
divergencies, calculate higher-derivative counterterms, and find a
recurrent formula for the local parts of the reaction force in
neighboring dimensions. Higher-dimensional curved space counterterms
are not simply the covariant generalizations of the flat ones, but
contain additional curvature-dependent terms. We illustrate our
formalism in four and six dimensions. In the first case we rederive
the results of DeWitt-Brehme-Hobbs in a simpler way, in the second
case we give a covariant generalization of the Kosyakov equation.
The local part of the reaction force is found to contain a term
proportional to the Riemann tensor which is absent in four
dimensions.
\end{abstract}
 \maketitle
\section{Introduction}

Though already Abraham \cite{Ab05} gave a correct formula for
radiation reaction force in classical electrodynamics
\begin{equation}\label{ab}
m \ddot z^\alpha= \frac{2}{3}e^2(\stackrel{...}{z}^{\alpha }-\ddot
z^{2}\dot z^{\alpha }),
\end{equation}
this problem, especially its generalization for curved space-time
and gravitational radiation still attract attention. Physical
status of the Eq. (\ref{ab})  also remains  subject of discussion.
In 1938 Dirac gave a consistent derivation of the Eq. (\ref{ab})
for a radiating charged which is now commonly referred as the
Lorentz-Dirac (LD) equation \cite{Dirac}.  There exist a wide
literature devoted to different aspects of this equation, its
solutions and applications (see
\cite{ivan,FuRo60,Ro61,Ba,Ro65,Te70,Po99,galpav,Ko07}, and refs.
therein). Physical interpretation of the LD equation is obscured
by the presence of the third derivative Schott term \cite{Sch15}
\begin{equation}
f^{\alpha}_{\mathrm{Schott}}=\frac{2}{3}e^2
\stackrel{...}{z}^\alpha.
\end{equation}
This term can be found \cite{LaLi}  as  improvement of the
radiation recoil term equal to the loss of the particle momentum
due to radiation,
\begin{equation}
f^{\alpha}_{\rm{emit}}=-\frac{2}{3}e^2 \ddot z^2 \dot z^\alpha,
\end{equation}
to ensure the orthogonality of the reaction force to the particle
four-velocity. But, somewhat unexpectedly, the Schott term leads
to disturbing modification of the particle momentum balance  and
creates the self-acceleration problem, for which reason the
validity of the LD equation is questioned \cite{LaLi}. A
widespread opinion is that the resolution of these problems
requires quantum theory.

Meanwhile, correct physical interpretation of the Schott term
solves this controversy within the classical theory.   As was
emphasized long ago by Teitelboim \cite{Te70}, and more recently
reconsidered (with all calculations explicit) in our paper
\cite{galpav}, the Schott term must be treated as a part of the
electromagnetic momentum, but not as a part of the particle
momentum. The LD equation (and its generalization which we will
discuss here) describes a composite object whose energy-momentum
contains the mechanical part and the Coulomb field part bound to
the charge, which varies once particle is accelerated. The
energy-momentum balance can  be verified globally for world-lines
which asymptotically are purely geodesic (non-accelerated). The
Coulomb coat varies with time when the charge experiences an
acceleration, moreover, as a result of interaction with the
charge, the coat energy can be transformed to that of radiation.
This provides a natural explanation of the Born paradox.

The LD equation was covariantly generalized to curved space-time
with an arbitrary metric in 1960 by DeWitt and Brehme
\cite{DeWitt:1960fc} and completed with the Ricci term by Hobbs
\cite{hobbs} (for a recent review see \cite{Poisson:2003nc}). In
this case, in addition to the local Abraham force (\ref{ab}),
there is a
 tail  integral over the past history of the
particle: \begin{align} \!\! m \ddot z^\alpha = e F^{\alpha
}_{\rm{ext}\beta } \dot z^\beta +
\frac{2}{3}e^2(\stackrel{...}{z}^{\alpha}-\ddot z^{2}\dot
z^{\alpha })+ \frac{e^2}{3} \left( R^{\alpha}_{\;\;\beta } \dot
z^\beta \right. \nnn  \!\!\!\!  + \!\!  \left. R_{\gamma\beta }
\dot z^\gamma \dot z^\beta \dot z^\alpha  \right) + e^2\, \dot
z^\beta \int\limits_{-\infty }^{\tau } f^{\alpha}_{\;\;\beta
\gamma }(\tau, \tau') \dot z^{\gamma }(\tau')d\tau '\! ,
\end{align} where $F^{\alpha }_{\rm{ext}\beta }$ is the external
electromagnetic field and $f^{\alpha}_{\;\;\beta \gamma } (\tau,
\tau')$ is some two-point tensor. It could be expected that the
curved space LD equation which takes into account non-local
effects, should violate the equivalence principle. But in the
DeWitt-Brehme-Hobbs (DWBH) equation this violation is minimal:
only the Ricci term is present in the local part of the LD force,
not the Riemann tensor. In empty space the local part of the DWBH
equation is just the covariantization of the flat-space LD
equation. The tail term of course makes the reaction problem
essentially non-local, but this term is due to the global
properties of space-time.

Similar equation was also derived for the scalar radiation
reaction \cite{quinn,quwa}: \begin{align} \label{quin} m(\tau
)\ddot z^{\alpha }= \frac{q^2}{3}
(\stackrel{...}{z}^{\alpha}-\ddot z^{2}\dot z^{\alpha })+
\frac{q^2}{6}\left(R^{\alpha}_{\;\;\beta } \dot z^\beta +\right.
\nnn \left.+ R_{\gamma\beta } \dot z^\gamma \dot z^\beta \dot
z^\alpha \right) -q^{2} \int^{\tau^{-} }_{-\infty }
G^{,\alpha}_{{\rm ret}}( z(\tau ),z(\tau' ) )d\tau ',
\end{align} where $G_{\rm{ret}}$ is the retarded solution of the wave equation and
 $m(\tau )=m_{0}-q\phi (z)$ is the variable effective mass.

The derivation of these generalizations of the LD equation to
curved space-time in the existing literature is based on lengthy
calculations of the integral contribution of the bound and
radiative energy-momentum fluxes within the world-tube surrounding
the particle world line. Meanwhile, the resulting equation
(including the tail term) depends only on the quantities {\em
localized on the world-line}. This indicates that a simpler
derivation should be possible which does not require considering
the  field outside the world line (in flat space such a
calculation is known for a long time \cite{ivan}). Recently we
have shown that similar derivation (restricted to four space-time
dimensions) can be performed in curved space-time as well
\cite{gaspist}.  Note that we discuss here only radiation of
non-gravitational nature, for gravitational waves the situation is
likely to be more complicated.

The purpose of the present paper is to generalize the approach of
\cite{gaspist} to higher-dimensional space-times.  Generalization
to higher dimensions in the flat space-time  was discussed
recently in
 \cite{Ko99,galtsov,kazinski1} in view of an interest to cosmological
models with large extra dimensions. Here we consider the curved
space of dimension $D\geqslant4$.
 Note that the cases of odd and
even dimensions are essentially different: in odd dimensions the
tail terms appear already in the flat space due to violation of
the Huygens' principle. (This effect can be easily interpreted
considering $D$-dimensional point particles as parallel strings in
$D+1$ dimensions \cite{galtsov}.)  We will present some general
considerations valid in any dimensions and then specialize to an
even-dimensional  case. The mostly plus space-time signature is
used.

\section{Green functions}\label{qq22}
As in $D=4$ \cite{DeWitt:1960fc}, the curved space Green's
functions for massless fields in arbitrary dimensions can be
constructed starting with the Hadamard solution.
\subsection{Hadamard expansion}\label{qq22sc}
We start with the scalar  Hadamard \cite{hada,decurw} Green's
function $G_H(x,x')$ which is a solution to  the homogeneous
scalar wave equation
\begin{equation} \label{phi1}
 \Box_x G_H(x,x') =0,
\end{equation}
where $\Box=g_{\mu\nu}\nabla^{\mu} \nabla^{\nu}$ is the curved
space scalar D'Alembert operator. The procedure consists in
expanding $G_H(x,x')$ in terms of the Synge world function
\vspace{-2mm}
\begin{equation}\label{syng}
    \sigma(x,x')= \frac12(s_1-s_0)\int\limits_{s_0}^{s_1}g_{\alpha\beta}{\dot
    z}^\alpha {\dot z}^\beta d\tau,
\end{equation}
  where an integral is taken along the geodesic
$x=z(\tau),\, z(\tau_0)=x,\, z(\tau_1)=x'$ connecting points $x$
and $x'$ and it is assumed that there is a unique such geodesic
(which makes the construction essentially local). The sign is
chosen such that $\sigma$ be negative for time-like geodesics. The
gradients of the Synge two-point function
\begin{equation}\label{gradsig}
 \sigma_\mu\equiv\partial\sigma(x,x')/\partial x^\mu,\quad
 \sigma_\alpha\equiv\partial\sigma(x,x')/\partial x'^\alpha
\end{equation}
satisfy the Hamilton-Jacobi equation
\begin{equation}\label{hamjac}
   g^{\mu\nu}\sigma_\mu \sigma_\nu=g^{\alpha\beta}\sigma_\alpha
   \sigma_\beta=2\sigma,
\end{equation}
where (using DeWitt-Brehme conventions) we denote by the initial
Greek letters $\alpha,\beta,\ldots$ tensor indices transforming at
$x'$, and by the older Greek letters $\mu,\nu,\ldots$ the indices
transforming at $x$. In the coincidence limit $x\to x'$  (denoted
by square brackets) one has
\begin{equation} \label{coilim}
[\sigma] \equiv\lim_{x\to x'} \! \sigma(x,x') \! =0,\;\;
[\sigma_{\alpha}]=0,\;\; [\sigma_{\mu}]=0.
\end{equation}
A mixed covariant derivative $\nabla_\alpha \nabla_\mu
\sigma(x,x')$, where according to the convention  $\nabla_\mu$
acts on $x$ and $\nabla_\alpha$ on $x'$, defines the van Vleck
determinant
\begin{equation}\label{vanvlek}
   \Delta(x,x')\equiv \frac{\det \left[\nabla_\alpha \nabla_\mu
\sigma(x,x')\right]}{\sqrt{g(x)g(x')}}.
\end{equation}
Higher covariant derivatives of the Synge function in the
coincidence limit are expressed via the metric tensor and the
Riemann tensor, this is used to find Taylor's expansions of the
two-point tensors \cite{DeWitt:1960fc}.

Another useful two-point tensor is the bivector of parallel
transport, which can be expressed through the vielbein associated
with metric: $g_{\mu\nu}(x)=e_\mu^m(x) e_\nu^n(x)\eta_{mn}$ as
\begin{equation}\label{bivepar}
{\bar g}_\mu^{\;\;\alpha}(x,x')=e_\mu^m(x)e^\alpha_a(x')
\delta^a_m.
\end{equation}
This operator transforms indices of the type $\alpha$ into indices
of the type $\mu$, in particular,
\begin{equation}\label{inde}
   {\bar g}_\mu^{\;\;\alpha}(x,x')\sigma_\alpha(x,x')=-\sigma_\mu(x,x'),
\end{equation}
where it was taken into account that the two gradients have
opposite directions.

 The world function has a dimension of the
length squared. In flat space-time
\begin{equation}\label{flasig}
   \sigma(x,x')= \frac{1}{2} (x-x')^2.
\end{equation}
A two-point {\em bitensor} (in particular, a biscalar) can be
expanded in powers of Synge's world function and its gradients. It
is worth noting that this type of expansion is  different from the
Taylor's covariant expansion of a two-point bitensor. In
particular, vanishing of $\sigma(x,x')$ does not necessarily mean
$x=x'$.

Let us write an appropriate expansion for the Hadamard's two-point
scalar function $G_H(x,x')$ which is singular in the coincidence
limit. For $D=4$ the Hadamard expansion contains two terms
singular in $\sigma$, namely, $\sigma^{-1}$ and $\ln \sigma$, the
remaining part being regular in $\sigma$. In higher dimensions one
has to add other singular terms, and by dimensionality it is easy
to guess that each dimension introduces   an additional factor
$\sigma^{-1/2}$. Thus, the Hadamard expansion in $D=2d$ dimensions
($d\geqslant 3/2$ is integer or half-integer) generically must
read
\begin{equation}\label{phi3} G_H(x,x') =\frac{1}{(2\pi)^{d}}
\left[\sum_{n=1}^{D}g_n \sigma^{1-n/2}+v \ln \sigma\right],
\end{equation}
where $g_n=g_n(x,x'),\; v=v(x,x')$ are coefficient two-point
scalar functions. The logarithmic term, as will be clear shortly,
is present only in even dimensions. Let us show, that in odd
dimensions we actually have only odd powers of $\sigma^{-1/2}$,
and in even dimensions --- only even powers, that is, an expansion
in terms of inverse integer powers of $\sigma$.

Substituting  (\ref{phi3}) into (\ref{phi1}), in the leading
singular order we will have:
\begin{equation} \label{phi4}
 2\sigma^\mu \partial_\mu g_D+g_D(\Box\sigma-D) =0.
\end{equation}
Using the relation
\begin{equation}\Box\sigma=D-\frac{\sigma^\mu\partial_\mu\Delta}{\Delta}
\end{equation} we arrive at
the equation
\begin{equation} \label{phi5}
 \sigma^\mu (2\,\partial_\mu g_D-g_D\partial_\mu \ln\Delta) =0,
\end{equation}
which is equivalent to
\begin{equation} \label{phi6}
 2 \, \partial_\mu g_D-g_D\partial_\mu \ln\Delta =\left(g^{\mu\nu}-
\frac{\sigma^{\mu}\sigma^{\nu}}{2\sigma} \right)h_{\nu},
\end{equation}
where $h_{\nu}(x)$ is an arbitrary regular vector field. The left
hand side of this equation is analytic in the coincidence limit,
while the right hand side is not, unless $h_{\nu}(x)=0$, so with
the normalization condition $[g_D=1]$ we obtain
\begin{equation} \label{phi7}
g_{D}=\Delta^{1/2}.
\end{equation}
In the next singular order   we obtain:
\begin{equation} \label{phi8}
  2 \, \partial_\mu g_{D-1} \sigma^\mu +g_{ D-1 }\Box\sigma -  (D-1)
g_{ D-1 }=0.
\end{equation}
For $D=3$, this equation holds not for the $g_{D-1}$, but for $v$.
The Eq. (\ref{phi8}) has a solution
\begin{equation} \label{phi9}
g_{D-1}=C\frac{\Delta^{1/2}}{\sigma^{1/4}},
\end{equation}
which does not satisfy the required analyticity  for $g_n$ unless
 $C=0$, and hence
\begin{equation} \label{phi9s}
g_{D-1}=0.
\end{equation}
For $D=3$ this means the absence of the logarithmic term.
Similarly, considering the equation for $g_{(D-1-2k)},\;k \in
\mathbb{N}$  we find
\begin{equation} \label{phi9u}
g_{D-1-2k}=0.
\end{equation}
This means that for an even-dimensional space-time the Hadamard
Green's function contains only integer negative powers of
 $\sigma$ plus logarithm and a regular part, while in the odd-dimensional case -- only
half-integer powers of $\sigma$ plus a regular part.

\subsection{ Dimensional recurrent relation } For the sequence of Green's
functions in the flat space-time, the one in  $D+2$ dimension is
proportional to the derivative  of the Green's function in the
twice preceding dimension $D$. In particular, in even dimensions
the symmetric Green's function is the derivative of the order
$d-2$ of the delta-function:
\begin{equation}G^{D}\sim \delta^{d-2}(-\sigma),\quad\sigma=(x-x')^2/2 \end{equation}
and thus, $G^{D+2}\sim d G^{D}/d\sigma$. Applying a convenient
regularization of the delta function,
\begin{equation}\delta((x-x')^2)=\lim_{\varepsilon \to
+0}\delta( |(x-x')^2|-\varepsilon^2),\end{equation} we obtain
\begin{equation}G^{D+2} \sim d G^{D}/d\varepsilon^2.\end{equation}
 This relation has a
consequence that the Laurent expansion of the Lorentz-Dirac force
in terms of $\varepsilon$ in the even-dimensional Minkowski space
has only odd negative powers, and no even terms.

Now consider the Hadamard function in the curved space. Taking
into account the Eq. (\ref{phi9u}), we change notation for $g_n$
in (\ref{phi3}) in even  dimensions as follows:
\begin{equation} \label{red3}
G_H=\frac{1}{(2\pi)^{d}}\left(\sum_{k=0}^{d-2}\frac{u_k}{\sigma^{d-1-k}}+
v \ln\sigma + w\right),
\end{equation}
where $u_0 =\Delta^{1/2}$ and we also denote $v= u_{d-1},\;w=
u_{d}$.  Applying the curved space operator  $\Box$  with respect
to $x^{\mu}$, we obtain from $\Box G_H=0$ the following equation:
\begin{align}\label{red2}
 & \frac{d-2}{\sigma^{d-1}}\left(\frac{\Box u_0}{d-2}
-2(\partial_\mu u_{1} -u_{1}\partial_\mu\ln u_0)
\sigma^{\mu}-\!\!2u_{1}\!\!\right) + \nnn
&\!\!+\frac{d-3}{\sigma^{d-2}}\left(\frac{\Box
u_1}{d-3}\!-\!2(\partial_\mu u_{2}  +u_{2}\partial_\mu\ln u_0
)\sigma^{\mu}\!\!-4u_{2}\right)+ \nnn
&...+\frac{1}{\sigma^2}\left(\Box u_{d-3}-2\partial_\mu
u_{d-2}\sigma^{\mu}+(4-\Box \sigma)u_{d-2}\right)+
 \nnn
&+\frac{1}{\sigma}\left(\Box u_{d-2}+2(\partial_\mu u_{d-1}
-2u_{d-1}\partial_\mu\ln u_0) \sigma^{\mu}+\right. \nnn
&\left.+(D-2)u_{d-1}+\sigma \Box u_{d}\right)+\ln{\sigma}\Box
u_{d-1}=0. \end{align}Each term in the right hand side must vanish
independently, which gives the system of recurrent differential
equations for $u_i (x,x')$ . Integrating them along the geodesic
connecting two points $x,x'$, one can uniquely extract $u_1(x,x')$
through $u_0(x,x')$. Furthermore, $u_2$ is determined through
$u_1$, etc.

Now we want to compare $u^{D}_i(x,x')$ in different dimensions
$D$. First of all we observe that $u_0= \Delta^{1/2}$ in any
dimension. It is worth noting, however, that in the expansion in
terms of $\sigma$
\begin{equation} \label{vVleckexp}
u_0^{D}=\Delta^{1/2}=1+1/12\,R_{\alpha\beta}\sigma^{\alpha}\sigma^{\beta}+...
\end{equation}
the tensor indices $\alpha,\,\beta$ will run different range of
values. Keeping this subtlety in mind, we can write
\begin{equation}u_{0}^{D'} = u_{0}^{D}\end{equation} for any $D,\,D'$.
Now, equating the first line in (\ref{red2}) to zero, we see that
the product $ u_{1}(d-2)$ is subject to the same master equation
in any dimension, so for $D,D' \geqslant 5$ one has
\begin{equation} \label{red2a}
u_{1}^{D'} = u_{1}^{D}\frac{d-2}{d'-2}.
\end{equation}
The second line in (\ref{red2} ) gives similarly for $D,D'
\geqslant 7$:
\begin{equation} \label{red2b}
u_{2}^{D'} =u_{2}^{D}\frac{d-2}{d'-2}\;\frac{d-3}{d'-3}.
\end{equation}
If $D'=D+2$, this relation can be rewritten as
\begin{equation} \label{red2b1}
u_{2}^{D+2} =u_{2}^{D}\frac{d-3}{d-1}.
\end{equation}
For $u_k^{D+2}$ we obtain
\begin{equation} \label{red2b1a}
u_{k}^{D+2} =u_{k}^{D}\frac{d-1-k}{d-1}.
\end{equation}

Comparing the equation for the $\sigma^{-1}$ term in  $D$
dimensions and the equations for the $\sigma^{-2}$ term in
$D'=D+2$ dimensions
\begin{equation} \label{red2b1b}
u^{D+2}_{d-1} =- u_{d-1}^{D}\frac{1}{d-1}=- v^{D}\frac{1}{d-1}.
\end{equation}
Denote the part of the Hadamard function (\ref{red3}) without the
logarithmic term as ``direct'' (it will correspond to the
non-scattered propagation of waves)
\begin{equation}
G_{H\,{\rm dir}}=\frac{1}{(2\pi)^{d}}
\sum_{k=0}^{d-2}\frac{u_k}{\sigma^{d-1-k}}.
\end{equation}
Differentiating $G_H $ with respect to $\sigma$ with the rule that
the functions $u_k, v, w$ are \textit{not} differentiated over
$\sigma$ we find
 \begin{align} \label{red4}  (2\pi)^{d}\frac{\partial G_H
}{\partial \sigma}=& -\sum_{k=0}^{d-2}\frac{u_k
}{\sigma^{d-k}}\left(d-1-k\right)+ \frac{v}{\sigma} =\nnn =&
-\sum_{k=0}^{d-2}\frac{u_k }{\sigma^{d-k}}\left(d-1-k\right).
\end{align}Taking into account the relation (\ref{red2b1b}), one obtains
\begin{align} \label{red5} \frac{-  (2\pi)^{d}}{d-1}\frac{\partial
G_H^{D}}{\partial \sigma}=&
\sum_{k=0}^{d-2}\frac{u_k^{D+2}}{\sigma^{d-k}}+
\frac{u_{d'-2}^{D'}}{\sigma}=\nnn =&
\sum_{k=0}^{d'-2}\frac{u_k^{D'}}{\sigma^{d'-k-1}}=
(2\pi)^{d'}G_H^{{\rm dir }\,D'}, \end{align}where $D'=D+2$.

\subsection{Retarded Green's functions}
To define the retarded and advanced Green's function one has to
specify boundary conditions. For a general space-time this is not
suggestive, so usually one uses a quasilocal definition proposed
by DeWitt and Brehme \cite{DeWitt:1960fc} in the case of four
dimensions. We follow the same approach here for arbitrary $D$.
The prescription consists in constructing the Feynman's propagator
$G_F$ by means of a shift of $\sigma$ in the Hadamard expansion
onto the  upper complex plane $\sigma\to\sigma+i0$ and then
separating real and imaginary parts
\begin{equation}\label{imre}
 G_{F}=G_{(1)}-2i  G_{\rm self}.
\end{equation}
This prescription works both for odd and even $D$, here we will
concentrate on $D$ even. In this case the relations
\begin{align}\label{6D2}\lal
(\sigma+i0)^{-n}=\mathcal{P}\frac{1}{\sigma^n}-\frac{(-1)^{n-1}
i\pi}{(n-1)!} \delta^{(n-1)}(\sigma) \\ \lal
\ln(\sigma+i0)=\ln|\sigma|+i\pi\theta(-\sigma) \end{align}are
useful and we find for the symmetric (self) Green's function:
\begin{align}\label{simmeD} \lal
 G_{\rm self} =
-\frac{1}{2}\mathrm{Im} G_{F}=\frac{1}{4(2\pi)^{d-1}}\times \nnn
\lal
 \times
 \left[\sum_{m=0}^{d-2}\frac{(-1)^{m} u_{d-2-m}
\delta^{(m)}(\sigma)}{m!}-v \theta (-\sigma )\right].
\end{align}To construct the retarded and advanced Green's
functions  one applies the formal global definition of the
Heaviside function $\theta [\Sigma (x),x']$ which is equal to one
if $x'$ is in the past of any space-like hypersurface $\Sigma (x)$
containing $x$, and zero otherwise.  Then the symmetric Green's
function is split into the sum of the retarded and advanced ones
like in the flat space:
\begin{equation} \label{ret}
G_{\rm {ret}}  (x,x')=2\theta [\Sigma (x),x'] G_{\rm {self}}
(x,x') ,\end{equation}
 \begin{equation} \label{adv}
G_{\rm {adv}}  (x,z) =2\theta [x',\Sigma (x)] G_{\rm {self}}
(x,x').
\end{equation}
The retarded and advanced Green's functions satisfy the
inhomogeneous wave equation:
\begin{equation}\label{hino}
 \Box G (x,x')= - \frac{\delta^{D}(x-x')}{g^{1/4} (x)g^{1/4}
(x') (d-2)!},
\end{equation}
where a symmetrization of the delta-density is performed.
Following the above prescription one finds the retarded Green's
function in terms of the same $u_k,\, v$ (note that the regular
term $w $ does not enter): \begin{align} \label{red5a} & G_{\rm
ret} =\frac{1}{2(2\pi)^{d-1}}\Theta(x',\Sigma(x))\times \nnn  &
\times \left( \sum_{m=0}^{d-2}\frac{(-1)^{m} u_{d-2-m}
\delta^{(m)}(\sigma)}{m!}- v \theta(- \sigma)\right). \end{align}
Here the sum contains terms proportional to derivatives of the
delta function constitute the direct part of the retarded function
with the support on the ``light cone'' $\sigma=0$:
\begin{equation} \label{red5l}
 G_{\rm dir}\! =\!\frac{1}{2(2\pi)^{d-1}}\Theta[\Sigma]\!\!
\sum_{m=0}^{d-2}\!\frac{(-1)^{m} u_{d-2-m}
\delta^{(m)}(\sigma)}{m!}.\!
\end{equation}
The $v$-term is localized inside the light cone and represents a
tail resulting from the scattering of waves on the curvature.

Using the recurrent relation found for the Hadamard function in
the previous section, we obtain for the retarded Green's function
a similar relation
\begin{equation} \label{red6}
\frac{\partial G_{\rm ret}^{D}}{\partial \sigma} =
-2\pi\left(d-1\right)G_{\rm dir}^{D+2}.
\end{equation}
This relation  can be rewritten using the  regularized
delta-functions in terms of differentiation with respect to the
point-splitting distance $\mathcal{E}=\varepsilon^2/2$,
\begin{equation} \label{delreg}\delta(-\sigma)=\lim_{\varepsilon \to
+0}\delta(|\sigma|-\mathcal{E}).\end{equation} Obviously we will
have
\begin{equation} \label{red7}
 G_{\rm dir}^{D+2}= \frac1{2\pi(1-d)} \frac{\partial G_{\rm
ret}^D}{\partial\mathcal{E}}.
\end{equation}
\subsection{Vector field}\label{qq22em} Now consider the massless vector
field $A_\mu$ in $D$ dimensions governed by the usual Maxwell
action quadratic in $F_{\mu\nu}=\nabla_\mu A_\nu-\nabla_\nu
A_\mu$. The homogeneous Maxwell equation
\begin{equation}\label{superstar}
F^{\mu \nu }_{\;\;\;\;; \,\nu }=0
\end{equation}
in terms of $A_\mu$ in the gauge  $A^{\mu }_{\;\;;\,\mu }=0$ reads
\begin{equation}\label{jpo}
\Box A_{\mu }-R_{\mu }^{\nu }A_{\nu }=0,
\end{equation}
where $R_{\mu }^{\nu }$ is the Ricci tensor (we use the definition
as $R^{\mu\nu }=g_{\rho\sigma}R^{\sigma \mu\rho \nu }$). Similarly
to the scalar case, we define the Hadamard function as a solution
to the equation
 \begin{equation}  \label{jpo1}
\Box_x G_{\mu \alpha}(x,x') -R_{\mu }^{\nu }(x)G_{\nu
\alpha}(x,x') =0.
\end{equation}
We look for a solution in the form ($D$ even) \begin{align}
\label{red1} G_{
H}^{\mu\alpha}=&\frac{1}{(2\pi)^{d}}\left(\frac{u_0^{\mu\alpha}}{\sigma^{d-1}}+
\frac{u_1^{\mu\alpha}}{\sigma^{d-2}}+\frac{u_2^{\mu\alpha}}{\sigma^{d-3}}+...+\right.
\nnn &+
\left.\frac{u_{d-3}^{\mu\alpha}}{\sigma^{2}}+\frac{u_{d-2}^{\mu\alpha}}{\sigma}+v^{\mu\alpha}\ln{\sigma}
+u_{d}^{\mu\alpha}\right), \end{align} where all the quantities
$u_k^{\mu\alpha}(x,x'),\,v^{\mu\alpha}(x,x') $ are two-point
tensors regular in the coincidence limit. Now the lowest order
term is
\begin{equation} \label{rew0}
u_{0}^{\mu\alpha}=\Delta^{1/2} {\bar{g}}^{\mu\alpha}.
\end{equation}
  By the same
reasoning as in the scalar case we find   relations between
$u_k^{\mu\alpha}$ in  different dimensions and  establish  the
differentiation rule essentially equivalent to (\ref{red7}).

\section{Radiation reaction} \subsection{Lorentz-Dirac force}
Consider a point particle of mass $m$ interacting with  massless
scalar and vector fields  ($q\,,e$ being the corresponding
couplings). The particle world-line will be denoted as
$x^\mu=z^{\mu}(s)$ where $s$ is the interval. The action consists
of the field part
\begin{equation}
S_F=-\frac{1}{\Omega_{D-2}}\int \left( (\nabla \phi)^2+\frac{1}{4}
F^2 \right) \sqrt{-g}d^Dx,
\end{equation}
and the particle term
\begin{equation}
S_p=- m_{0}\int  (1+q\phi)\sqrt{-\dot{z}^2}d\tau -e\int
A_{\mu}\dot{z}^{\mu}d\tau,
\end{equation}
where  $\phi$ and $A^{\mu}$ are taken at the point $x=z(\tau)$.
The field equations read \begin{align}\label{emotions1} \lal \Box
\phi =\Omega_{D-2}
 \rho   \\ \lal \label{emotions2}F^{\mu\nu}_{\quad;\nu}= \Omega_{D-2}
j^{\,\mu}, \end{align}while the equation of motion of a particle
is
\begin{equation}
\label{emotions3}m_0(1+q\phi)\ddot{z}^{\mu}=-eF^{\mu}_{\;\;\nu}\dot{z}^{\nu}-m_0
q\Pi^{\mu\nu}\phi_\nu,
\end{equation}
 where the covariant differentiation along the world line is denoted by dot,
 $\phi_\nu=\partial_\nu\phi$, and
$\Pi^{\mu\nu}=g^{\mu\nu}+\dot{z}^{\mu}\dot{z}^{\nu}$ is a
projector on the subspace orthogonal to the world line. The
natural parameter $\tau$  is assumed, so that $\dot{z}^2=-1$.

 To get the radiation
reaction force we insert at the right hand side of the Eq.
(\ref{emotions3}) the retarded solution of the wave equations
(\ref{emotions1},\ref{emotions2})
 \begin{align}\label{retscal} \lal
\phi(x)=\int G_{\rm ret}(x,x')\rho(x')\sqrt{-g(x')}dx', \\
 \lal
\label{retvect}A^\mu(x)=\int G_{\rm
ret}^{\mu\alpha}(x,x')j_{\alpha}(x')\sqrt{-g(x')}dx', \end{align}
with the sources \begin{align}\label{curscal} \lal \rho(x') =m_0 q
\int \frac{\delta
(x'-z(\tau))}{\sqrt{-g}}d\tau, \\
 \label{curvect} \lal
j^\mu(x')=e\int \frac{\delta
(x'-z(\tau))\bar{g}^{\mu}_{\,\alpha}\dot{z}^{\alpha}(\tau)}{\sqrt{-g}}d\tau.
\end{align} The resulting Lorentz-Dirac self-force will be by definition
the right hand side of the  equation of motion in absence of
external fields
\begin{equation} m_0 \ddot{z}^{\mu}=f^{\mu}_{\rm
LD},\end{equation}
 where the mass parameter at the left hand side is
constant (contrary to another frequently used in presence of the
scalar field definition of the effective mass in (\ref{quin})).
The Lorentz-Dirac force consists of a scalar part
\begin{equation}\label{sc}
 f^{\mu}_{\rm sc}  =-m_0q\Pi^{\mu\nu}(\phi_{\nu}+\phi
\ddot{z}_{\nu}),
\end{equation}
and a vector part
\begin{equation} \label{em}
f^{\mu}_{\rm em}=-eF^{\mu}_{\;\;\nu}\dot{z}^\nu.
\end{equation}

According to the above decomposition of the retarded Green's
function into the direct part and the tail part, we will have a
similar decomposition for the Lorentz-Dirac force. Then for the
direct part we obtain the following differentiation rule relating
the values of the force in twice neighboring dimensions:
\begin{equation}\label{red7a}
f^{\mu \; {\rm dir}}_{D+2}=-\frac{1}{D-1}\frac{\partial f^{\mu \;
{\rm dir}}_{D}(s,\mathcal{E})}{\partial \mathcal{E}}.
\end{equation}
The limit $\mathcal{E} \to +0$ has to be taken after the
differentiation. The direct force is due to the light cone part of
the retarded Green's function. This is not the full local
contribution to the Lorentz-Dirac force. An additional
contribution comes from the differentiation of the theta function
in the tail term $v  \, \theta(-\sigma)$. In the scalar case this
contribution vanishes, but in the electromagnetic case an extra
local term arises:
\begin{equation} \label{rew2}
f^\mu_{\rm loc}=e^2\left([v_{\mu\alpha}]\dot{z}_{\nu}
-[v_{\nu\alpha}]\dot{z}_{\mu}\right)\dot{z}^{\nu}\dot{z}^{\alpha},
\end{equation}
where the coincidence limit  $[v_{\nu\alpha}]$ depends on the
dimension. In the cases for $D=4$  it is given by (\ref{orto}).
The remaining contribution from the tail term will have the form
of an integral along the past half of the particle world line.
\subsection{Divergences}
The direct part of the Lorentz-Dirac force contains divergences.
To separate the divergent terms one can use the decomposition of
the retarded potential suggested in the case of four dimensions by
Detweiler and Whiting \cite{Detw02}. In higher even dimensions we
can follow essentially the same procedure. We define the
``singular'' part $G_{\rm S} $ of the retarded Green's function as
the sum of the symmetric part (``self'') and the tail function $v$
as follows \begin{align} \label{det0} G_{\rm S}(x,x')  = G_{\rm
self}(x,x')+\frac{1}{4(2\pi)^{d-1}}v(x,x')=\nnn =G_{\rm
self\;dir}(x,x')+\frac{1}{4(2\pi)^{d-1}} v(x,x')\, \theta(\sigma).
\end{align} Here the direct part of the self function means its part
without the tail $v$-term.
 Taking into
account $\Box v=0$, it is clear that $G_{\rm S}$ satisfies the
same inhomogeneous equation as $G_{\rm self}$. The $v$-term in the
second line of the Eq. (\ref{det0}) is localized outside the light
cone. Therefore the corresponding field (for instance, scalar), at
an arbitrary point $x$ will be given by
\begin{equation} \label{det1}
\phi_{\rm S} (x) =\phi_{\rm self\;dir}+\frac{m_0 q
\Omega}{4(2\pi)^{d-1}} \int \limits_{\tau_{\rm ret}}^{\tau_{\rm
adv}} v(x,z(\tau))d\tau,
\end{equation}
where the retarded and advanced proper time values $\tau_{\rm
ret}(x),\, \tau_{\rm adv}(x)$ are the intersection points of the
past and future light cones centered at $x$ with the world line.

The remaining part of the Green's function
\begin{equation}G_{\rm
R}(x,x')=G_{\rm ret}(x,x')-G_{\rm S}(x,x')
\end{equation}
 satisfies a
free wave equation and is regular.

Let us investigate divergent terms in the Lorentz-Dirac force and
show  that they can be eliminated adding counterterms into the
action (in flat space this was demonstrated for $D=6$ by Kosyakov
\cite{Ko99}  an generalized to arbitrary $D$ in \cite{kazinski2}).
Inserting (\ref{det1}) into the Lorentz-Dirac force defined on the
world-line $x=z(\tau)$, we observe that the integral contribution
from the tail term   vanishes and so the divergent part is
entirely given by  $\phi_{\rm self\;dir}.$ For the delta-functions
we use the point-splitting regularization (\ref{delreg}). All
divergent terms will then arise as negative powers of
$\varepsilon$. To prove the existence of the counterterms we
consider the interaction term in the action substituting the field
as the S-part of the retarded solution to the wave equation. In
the scalar case we will have:
\begin{equation} \!\! S_{\rm S}\!=\!\frac{1}2 \int G_{\rm
S}(x,x')\rho(x)\rho(x')\sqrt{g(z)g(z')}\,dx \,dx',
\end{equation}
where a factor one half is introduced to avoid double counting
when self-interaction is considered. Substituting the currents we
get
 the  integral over $d\tau,d\tau'$:
\begin{equation} \label{det2}
S_{\rm S}=\frac12m_0^2 q^2 \Omega \int G_{\rm
S}(z(\tau),z(\tau'))d\tau d\tau'.
\end{equation}
Since the Green's function is localized on the light cone (by
virtue of (\ref{det1})), we can expand the integrand in terms of
the difference $t=\tau-\tau'$ around the   point $z(\tau)$:
\begin{equation} \label{det3}
S_{\rm S}\sim \int d\tau \int \sum_{k,l}
B_{kl}(\tau)\delta^{(k)}(t^2-\varepsilon^2)t^l dt.
\end{equation}
Here  the coefficients $B_{kl}(z)$ depend on the curvature, while
the delta-functions are flat: $\delta^{(k)}(t^2-\varepsilon^2)$
(the derivatives with respect to the argument are understood). By
virtue of parity, the  integrals with odd $l$ vanish, so only the
odd inverse powers of $\varepsilon$ will be present in the
expansion. Moreover, if we know the divergent terms in some
dimension $D$, we can obtain all  divergent terms in $D+2$ except
for $1/\varepsilon$ term.  The linearly divergent term
corresponds to $l=2k$. The integral is equal to
\begin{equation}
\int\limits_0^{\infty} \delta^{(k)}(t^2-\varepsilon^2)t^{2k}
dt=\frac{(-1)^{k}(2k-1)!!}{2^{k+1}}\frac{1}{\varepsilon}.
\nonumber
\end{equation}
In four dimensions this term is unique. Applying our recurrence
chain  we obtain the inverse cubic divergence in six dimensions
and calculate again the linearly divergent term. Thus in $D=2d$
dimensions we will get $d-1$ divergent terms from which $d-2$ can
be obtained by the differentiation of the previous-dimensional
divergence, and the linearly divergent will be new. This linearly
divergent self-action term in the action will have generically the
form
$$
S_{\rm S}^{(-1)}\sim \frac{1}{\varepsilon}\int \sum_{k=n-2}^{2n-4}
\frac{(-1)^{k}(2k-1)!!}{2^{k+1}} B_{k,2k}(\tau)  d\tau. $$ Here
the coefficient functions are obtained taking the coincidence
limits of the two-point tensors involved in the expansion of the
Hadamard solution. They actually depend on the derivatives of the
world line embedding function $z(\tau)$ as well as the curvature
terms taken on the world line:
$$
B_{k,2k}(\tau)=B_{k,2k}(\dot{z},\ddot{z} ,..., R(z(\tau)),
R_{\mu\nu}(z(\tau)),...).
$$
 Finally we have to
rewrite the expression   in the curved reparametrization-invariant
form. This amounts to the replacement \begin{align}
 & \dot{z}(\tau) \to Dz\equiv
 \frac{1}{\sqrt{-\dot{z}^2}}\frac{dz}{d\tau}\nnn
& \ddot{z}(\tau) \to D^2 z, \quad \ldots \; ,\nnn  & d\tau \to
\sqrt{-\dot{z}^2} d\tau, \end{align} where $D$ is the covariant
derivative along the world line.

The vector case is technically the same, now one has to  expand in
in powers $t$ in the integrand of
$$S^{\rm S}=\frac12\int G^{\rm
S}_{\mu\alpha}(x,x')j^{\mu}(x)j^{\alpha}(x')\sqrt{g(z)g(z')}\,dx
\,dx'.$$

From this analysis it follows that in any dimension the highest
divergent term can be absorbed by the renormalization of the mass
as in the generating four dimensional case. To absorb the
remaining $d-2$ divergences one has to add to the initial action
the sum of $d-2$ counterterms depending on higher derivatives of
the particle velocity.
\subsection{Finite reaction terms}
To obtain local part of the finite radiation reaction terms one
can used the recurrent procedure from $D$ to $D+2$ dimensions. So
if one performs expansions in terms of $t=\tau-\tau'$ in the
integrals up to sufficiently high order, one can find the
generating expression to get finite reaction terms in higher
dimensions. In addition one obtains the integral tail term which
is essentially the same (in terms of function $v$) in all
dimensions.

\section{Four dimensions}\label{4dim}

Here we apply our formalism to rederive  the radiation reaction
force in four dimensions. This approach is much simpler than the
non-local DeWitt-Brehme type calculations (integration of the
stress-tensor flux over the world-tube surrounding the particle)
and gives a more transparent understanding of the renormalization
involved. Brief account of our approach in four dimensions was
presented in \cite{gaspist}. Calculation in four dimensions may
also serve as a starting point for higher-dimensional problems
using the recurrent relations found above.
\subsection{Scalar force}
In four dimensions the scalar retarded Green's function contains a
single direct term localized on the light cone  and a tail term:
\begin{equation}\label{4ret}
 G_{\rm ret} = \frac{1}{4\pi} \theta [\Sigma (x),x']
 \left[\Delta ^{1/2} \delta (\sigma )-v \theta (-\sigma )\right].
\end{equation}
The retarded solution for the scalar field reads
\begin{equation} \label{scagree}
 \phi_{\rm ret}(x)=m_0 q \!\!\int\limits_{-\infty}^{\tau_{\rm
ret}(x)}\!\!\left[ -\Delta ^{1/2} \delta (\sigma)+v \theta
(-\sigma )\right] d\tau'.
\end{equation}
Differentiating this expression we obtain
%Substituting the retarded potential (\ref{scagree}) into the right
%hand side of the equation (\ref{sc}), we obtain for
$\phi_{\nu}$ on the world line:
 \begin{align} \label{forscal}
 \phi_{\nu}(z(\tau)) =  & m_0 q \int\limits_{-\infty}^{\tau} [-\Delta
^{1/2}\delta' (\sigma ) \sigma_{\nu }-\nnn & -\Delta
^{1/2}_{;\nu}\delta (\sigma )-v \delta(\sigma )\sigma _{\nu
}+v_{\nu}]\, d\tau ', \end{align} where integration is performed
along the past history of the particle. All the two-point
functions are taken on the world-line at the points
 $x=z(\tau)$ (``observation'' point) and
 $z'=z(\tau')$ ( ``emission'' point).

To compute  local contributions  from the terms proportional to
delta-functions and its derivative, it is enough to expand the
integrand in terms of  the difference $s=\Delta \tau=\tau -\tau '$
around the point $z(\tau)$.  The Taylor (covariant) expansion of
the fundamental biscalar $\sigma (z(\tau),z(\tau' ))$ is given by
\cite{christ76}:
\begin{equation}
 \sigma (z(\tau),z(\tau' ))= \sum_{k=0}^{\infty} \frac{1}{k!}D^k\sigma (\tau,\tau)
 (\tau-\tau')^k,
\end{equation}
where $D$ is a covariant derivative along the world-line:
$$
D\sigma\equiv\dot \sigma = \sigma _{\alpha} \dot z^\alpha,
\;\;\;\;D^2\sigma\equiv\ddot \sigma = \sigma _{\alpha\beta } \dot
z^\alpha \dot z^\beta +
  \sigma _{\alpha}\ddot z^\alpha,
$$
etc. Such an expansion exists since the difference $s=\tau -\tau'$
is a two-point scalar itself: this is the   integral from the
scalar function $\int (- \dot z^{2})^{1/2} ds,$ along the
world-line from $z(\tau)$ to $z(\tau')$. Taking the limits and
using   $\dot z^2(\tau)=-1$, we find:
\begin{equation}
\sigma (s)= - \frac{s ^2}{2}- \ddot z^2(\tau) \frac{s
^4}{24}+\mathcal{O}(s^5).
\end{equation}
To obtain an expansion of the derivative of $\sigma$ over $z^\mu
(\tau)$ one can expand $\sigma^ {\mu}(\tau, \tau -s)$ in powers of
$s$. This quantity transforms as a vector at  $z(\tau)$ and a
scalar at $ z(\tau ')$, this
\begin{equation}\label{signu}
\sigma ^{\mu}(s)=  s \left(\dot z^\mu -\ddot z^{\mu}\frac{s}{2}
+\stackrel{...}{z}^\mu\frac{s^2}{6} \right)+\mathcal{O}(s^4),
\end{equation}
where the index $\mu$ corresponds to the point $ z(\tau ) $:
$\sigma ^{\mu}=\partial \sigma (z,z')/\partial z_{\mu}$. Recall
that the initial
 Greek indices  correspond to $z(\tau')$. The expansion of
$\delta(-\sigma)$ will read:
\begin{equation}
\delta(-\sigma)=\delta(s^2/2)+s^4\frac{\ddot z^2(\tau)}{24}
\delta'(s^2/2)+...
\end{equation}
 where the derivative of the delta-function is taken with respect to
the full argument. Since the most singular term is $\Delta
^{1/2}\delta' (\sigma ) \sigma_{\nu }$, the maximal order giving
the non-zero result after the integration is $s^3$. (Note, that in
order to use our dimensional recurrent relations to obtain a
reaction force in higher dimensions we should perform an expansion
up to higher orders in $s$.) Thus, with the required accuracy,
$\delta(-\sigma)=2\delta(s ^2),$ and all the integrals for the
delta-derivatives are the same as in the flat space-time. This
allows us to use the same regularization of the  delta-functions
with double roots $\delta(s ^2)$ by the point-splitting. Expanding
the biscalar $\Delta^{1/2}$ and its gradient at  $z$ we have:
\begin{equation}\label{utochka}
\Delta^{1/2}= 1+\frac{s^2}{12}R_{\sigma  \tau }\dot z^\tau \dot
z^\sigma, \;\; \partial_\nu\Delta^{1/2}= \frac{s}{6} R_{\nu \tau
}\dot z^\tau.
\end{equation}
Combining all the contributions we obtain finally for the field
strength: \begin{align} \label{scalderiv} \phi_{\nu} =  m_0 q
\Biggl(& \frac {1}{2 \varepsilon }\ddot z_{\nu} -\frac {1}{3}
\stackrel{...}{z}_{\nu}-\frac {1}{6} R_{\nu \tau }\dot z^\tau
+\frac{1}{12}R \dot z_{\nu} -\nnn &- \frac{1}{6} R_{\gamma \delta
}\dot z^\gamma \dot z^\delta \dot z_\nu\ + \int
\limits_{\infty}^{\tau} v_{\nu}\, d\tau ' \Biggr), \end{align}
where the first term is divergent. Note that the terms $- 1/6
\,R_{\lambda\rho}\dot z^{\lambda} \dot z^{\rho} \dot z_\nu\ +1/12
\,R \dot z_{\nu}$ are annihilated by the projector $\Pi^{\mu\nu}$.
The retarded field $\phi$ itself is also singular on the
world-line:
\begin{equation}\label{scaltraj}
\phi (z)  =-\frac{m_0q}{\varepsilon} + m_0 q \int \limits_{-\infty
}^{\tau } v d\tau' .
\end{equation}

Collecting all the orders (\ref{scalderiv}-\ref{scaltraj}), we
obtain the following expression for the scalar part of the
self-force (\ref{sc}): \begin{align} \label{scfin} f^{\mu}_{\rm
sc}= m_0^2 q^2 \Biggl[ & \Pi^{\mu\nu} \Biggl(\frac {1}{3}
\stackrel{...}{z}_{\nu}+\frac {1}{6} R_{\nu \tau }\dot z^\tau -
\int \limits_{-\infty }^{\tau } v_{\nu}\, d\tau ' \Biggr) \nnn
 & +\ddot z^{\mu} \Biggl( \frac {1}{2
\varepsilon }-\int \limits_{-\infty }^{\tau } v d\tau' \Biggr)
\Biggr].
\end{align}\subsection{Vector force} Now consider the vector
contribution. The Hadamard function in four dimensions is
\begin{equation}  \label{luthor}
G_{H\,\mu \alpha }= \frac{1}{(2\pi)^2} \left ( \frac{u_{\mu \alpha
} }{\sigma }  + v_{\mu \alpha } \ln \sigma  + w_{\mu \alpha
}\right ),
\end{equation}
where $ u_{\mu \alpha }(x,x')$, $ v_{\mu \alpha }(x,x') $ and
$w_{\mu \alpha } (x,x')$ are bivectors. One find $u_{\mu\alpha
}=\bar g_{\mu \alpha }\Delta ^{1/2}.$ We look for power expansions
\begin{equation}\label{xxx}
v_{\mu \alpha}=\sum_{n=0}^{\infty}v^{(n)}_{\mu \alpha} \sigma^n,
\qquad w_{\mu \alpha}=\sum_{n=0}^{\infty}w^{(n)}_{\mu \alpha}
\sigma^n,
\end{equation}
we find the following equation for  $v^{(0)}_{\mu \alpha }$:
\begin{align} 2v^{(0)}_{\mu \alpha }+(2v^{(0)}_{ \mu \alpha \;;\nu}- &  v
^{(0)}_{ \mu \alpha }\Delta^{-1}\Delta_{;\mu }) \sigma^{;\nu}=\nnn
= &- \Box u_{\mu \alpha}+ R_{\mu }^{\;\;\,\nu }u_{\nu \alpha }
\label{lowe}. \end{align} Substituting  the covariant derivative
of the two-point tensors, we obtain the following coincidence
limit for $v_{\mu \alpha }$:
\begin{equation}\label{orto}
[v^{(0)}_{\mu \alpha }]=[v_{\mu \alpha }]= \frac{1}{2}\bar g_{\mu
}^{\;\;\beta }\left(R_{\alpha \beta }-\frac{1}{6} g_{\alpha \beta
}R\right).
\end{equation}
 The symmetric Green's function is given by
\begin{equation}\label{simme}
 G^{\rm {self}}_{\mu \alpha} = \frac{1}{8\pi }\left[u_{\mu \alpha}
 \delta (\sigma )-v_{\mu \alpha} \theta (-\sigma )\right],
\end{equation}
while the retarded one is
\begin{equation} \label{ret1a}
G_{\rm {ret}}(x,x')=2\theta [\Sigma] G_{\rm self} (x,x').
\end{equation}
The retarded vector-potential is \begin{align} \label{emgree}
 A^{\rm ret}_{\mu}(x)=&-4\pi e \int G^{\rm
ret}_{\mu\alpha}(x,z(\tau'))\dot{z}^{\alpha} d\tau'=\nnn  = & e
\!\! \int \limits_{-\infty}^{s_{\rm ret}(x)} \!\! \left[-u_{\mu
\alpha} \delta (\sigma )+v_{\mu \alpha} \theta (-\sigma )\right]
\dot{z}^{\alpha}d\tau'. \end{align} The field strength on the
world-line $x=z(s)$ will be \begin{align} \label{emintens}
 \! F^{\rm ret}_{\mu\nu}(z(s))\! =\! e \! \!\int \limits_{-\infty}^{s} \! \!\biggl( & u_{\mu
\alpha; \,\nu} \delta (\sigma )+u_{\mu \alpha} \sigma_{\nu}\delta
'(\sigma )+ v_{\nu \alpha ; \, \mu}+\nnn  + & v_{\mu \alpha}
\sigma_{\nu}\delta (\sigma )- \!\{\mu\longleftrightarrow
\!\nu\}\!\biggr)\dot{z}^{\alpha}d\tau', \!\end{align} so the
Lorentz-Dirac force will read
\begin{equation} \label{emintens1}
f^{\mu}_{\rm em}(s)=-e F^{\,\mu}_{\;\;\nu}(z(s))\dot{z}^{\nu}(s).
\end{equation}
Substituting all the expansions \begin{align} & \bar g_{\nu \alpha
} ( \tau, \tau -s)  \dot z^\nu (\tau )\dot
z^\alpha(\tau-s)=-1-\ddot z^2 \frac{s^2}{2} + O(s^3) \nnn  &
u_{\nu \alpha }^{\;\;\;\; ; \, \mu }\dot z^{\nu }(\tau )\dot
z^{\alpha }(\tau -s)= -\frac{s}{6}R^{\, \mu }_{\;\;\tau }\dot
z^{\tau } + O(s^2) \nnn & u^{\mu }_{\,\,\,\alpha;\nu}\dot z^{\nu
}(\tau ) \dot z^{\alpha } (\tau -s)= \frac{s}{6} R_{\nu \tau }\dot
z^{\nu } \dot z^{\tau }\dot z^{\mu } + O(s^2)\nnn & u_{\nu \alpha
}\sigma ^{;\mu }\dot z^{\nu }(\tau ) \dot z^{\alpha } (\tau-s )=
-s \dot z^\mu  + \frac{s^2}{2} \ddot z^\mu+\nnn & \;\;+ s^3 \left(
-\frac{1}{6} \stackrel{...}{z}^\mu -\frac{1}{12}R_{\lambda \nu}
\dot z^\lambda \dot z^\nu \dot z^\mu -\frac{1}{2} \ddot {z}^2 \dot
z^\mu \right) + O(s^4) \nn
 \end{align}
\begin{align}   & u^{\mu} _{\;\;\alpha }\sigma _{;\nu } \dot z^{\nu }(\tau
) \dot z^{\alpha } (\tau-s )= -s\dot z^\mu+\ddot z^\mu s ^2  \nnn
& \; \; +s^3 \left( -\frac{1}{6}\ddot z^2 \dot z^\mu -
\frac{1}{12}R_{\lambda \nu } \dot z^\lambda \dot z^\nu \dot z^\mu
-\frac{1}{2} \stackrel{...}{z}^\mu \right) + O(s^4)\nnn &  v_{\nu
\alpha }\sigma ^{\,\mu } \dot z^{\nu } \dot z^{\alpha }
 = \frac{s}{2}
 \dot z^\mu R_{\alpha  \nu } \dot z^\alpha \dot z^\nu +\frac{s}{12} R\dot  z^\mu + O(s^2)
\nnn & v^{\mu} _{\;\;\;\alpha }\sigma _{;\nu }\dot z^{\nu } \dot
z^{\alpha } = -\frac{s}{2}R^{\mu }_{\alpha }\dot z^\alpha  +
\frac{s}{12}R \dot z^\mu + O(s ^2), \end{align}
 after integration we find
\begin{align} \label{emfin} f^{\mu}_{\rm em}=
e^2\Biggl[&-\frac{1}{2\varepsilon}\ddot{z}^{\mu}+\Pi^{\mu\nu}
\left(\frac{2}{3}\stackrel{...}{z}_{\nu}+\frac{1}{3}R_{\nu\alpha}\dot
z^\alpha\right) + \nnn &
 +\dot z^{\nu }(\tau ) \!\!\int
\limits_{-\infty }^{\tau } \!(v^{\mu }_{\,\,\,\alpha;\nu} \!-
v_{\nu \alpha }^{\quad;\mu })\dot z^{\alpha } (\tau'
)\,d\tau'\Biggr].
\end{align}

\subsection{Renormalization and total force} Combining
the divergent parts of the expressions  (\ref{emfin}) and
(\ref{scaltraj}) into the unique mass-renormalization term, we get
\begin{equation}\label{cancell}
 m-m_0=\frac{1}{2\varepsilon}(e^2-m_0^2 q^2).
\end{equation}
Clearly, this   is just the flat-space result. Note that under the
condition:
\begin{equation}\label{cancell1}
m_0 |q|=|e|,
\end{equation}
the model is free from singularities and does not require
mass-renormalization. The final form of curved space Lorentz-Dirac
equation in this case will read: \begin{align}\label{emscfin} \!\!
m(\tau)\ddot{z}^{\mu} \!= e^2 & \Biggl[\Pi^{\mu\nu}
\Biggl(\stackrel{...}{z}_{\nu}+\frac{1}{2}R_{\nu\alpha}\dot
z^\alpha - \int \limits_{-\infty }^{\tau } v_{\nu}\, d\tau
'\Biggr)\nnn &\!+ \! \dot z^{\nu }(\tau )\!\!\!\int
\limits_{-\infty }^{\tau }\!\!(v^{\mu }_{\,\,\,\alpha;\,\nu} \!\!-
\!v_{\nu \alpha }^{\quad;\mu })\dot z^{\alpha } (\tau' )\,d\tau'
\! \Biggr],
\end{align}
where $$m(\tau)=m+\int \limits_{-\infty }^{\tau } v
d\tau'$$ is the $\tau$-dependent ``mass''.

This result coincides with the sum of the DeWitt-Brehme-Hobbs  and
the corresponding scalar equations \cite{quinn} which was
previously obtained within the DeWitt-Brehme non-local approach.

\section{Six dimensions}\label{6Dcurve}
The next even-dimensional case is $D=6$. The vector reaction force
in the flat space-time was previously given by Kosyakov
\cite{Ko99}. Here we give the curved space treatment and also add
the scalar field.
\subsection{Scalar force}\label{6Dcurvesc}
Using the general formalism of Sec. \ref{qq22} we can present the
Hadamard Green's function of the scalar field in six dimensions as
follows
\begin{equation} \label{6D1}
G_H=\frac{1}{(2\pi)^3}\left(\frac{\Delta^{1/2}}{\sigma^2}+\frac{u}{\sigma}+v\ln
\sigma+w\right).
\end{equation}
The corresponding retarded Green's function will be
\begin{equation} \label{6D3}
G_{\rm
ret}\!\!=-\frac{\theta(\Sigma)}{8\pi^2}\!\!\left(\!\Delta^{1/2}\delta'(\sigma)-u\delta(\sigma)+
v\theta(-\sigma)\right)\!\!,
\end{equation}
so that the  retarded solution is
\begin{equation} \label{6D3aaa}
  \phi_{\rm ret}(x)  \!\! =\!\!
  \frac{1}{3}  \!\! \!  \int\limits_{-\infty}^{\tau_{\rm {ret}}(x)}
\!\!\!\!\!\left(\Delta^{1/2}\delta'(\sigma)\!-\!u\delta(\sigma)
\!+\!v\theta(-\sigma)\!\right)d\tau'\!\!,
\end{equation}
where $\sigma=\sigma(x,z(\tau'))$.

We have to expand the two-point scalars $\Delta^{1/2}(x,z)$,
$u(x,z)$, $v(x,z)$ around $x$ and then set $x=z(\tau)$,
$z=z(\tau').$ To get all the local terms explicitly  it is
sufficient to expand $u(x,z)$ up to the second order in
$\sigma^{\mu}$ keeping only the leading term (coincidence limit)
for $v(x,z).$
%Для нее определяющие уравнения
The defining equations are: \begin{align}\label{6D4a}\lal
\Box\Delta^{1/2}-2u_{\mu}\sigma^{\mu}-
u\left(2-\frac{\Delta_{\mu}\sigma^{\mu}}{\Delta}\right)=0,
\\ \lal \Box
u+2v_{\mu}\sigma^{\mu}+
v\left(4-\frac{\Delta_{\mu}\sigma^{\mu}}{\Delta}\right)+\sigma \Box w=0, \label{6D4bsc}\\
\lal \Box v=0. \label{6D4csc} \end{align} Note that the
representation of the Hadamard function in inverse powers  of
$\sigma$ is non-unique, but this amounts only to a redefinition of
the regular part  $w$ which does not contribute to the retarded
solution
\begin{equation}\phi_{\nu}^{\rm
ret}(x)=-\frac{8\pi^2}{3}mq\int G_{;\,\nu}^{\rm
ret}(x,x')j(x')dx',\end{equation} where
 \begin{align}
\label{6D11ww} G_{\rm ret\,,\nu}=&\frac{1}{8
\pi^2}\biggl[\Delta^{1/2}\delta''(\sigma)\sigma_{\nu}+\Delta^{1/2}_{;\,\nu}\delta'(\sigma)
-u\delta'(\sigma)\sigma_{\nu}\nnn &
-u_{;\,\nu}\delta(\sigma)-v\delta(-\sigma)\sigma_{\nu}+
v_{\nu}\theta(-\sigma)\biggr].\!
\end{align} The subsequent calculations
consist in expanding two-point functions into powers of $\sigma$,
which are now needed up to higher orders than in the
four-dimensional case since the Green's function has higher degree
of singularity. In particular, we get
 \begin{align}\label{6D5qqq}
\Delta^{1/2}=&
1+\frac{1}{12}R_{\alpha\beta}\sigma^{\alpha}\sigma^{\beta}-
\frac{1}{24}R_{\alpha\beta; \,
\gamma}\sigma^{\alpha}\sigma^{\beta}\sigma^{\gamma}+\nnn &
+\left(\frac{1}{288}R_{\alpha\beta}R_{\gamma\delta}+
\frac{1}{360}R^{\rho\;\;\tau}_{\;\;\alpha\;\;\beta}R_{\rho\gamma\tau\delta}\right.+\nnn
& \qquad \left.+ \frac{1}{60}R_{\alpha\beta; \,
\gamma\delta}\right)
\sigma^{\alpha}\sigma^{\beta}\sigma^{\gamma}\sigma^{\delta}.
\end{align}Also, an expansion of the argument of delta-functions
will be \begin{align} &
\sigma=\frac{1}{2}\sigma^{\mu}\sigma_{\mu}=\frac{1}{2}\left(-\tau^2-\frac{1}{12}\ddot{z}^2
\tau^4+\frac{1}{12}(\ddot{z}\stackrel{...}{z})\tau^5\right),\nnn &
\delta(\sigma)=\delta(-\tau^2/2)+\delta'(-\tau^2/2)
\left(-\frac{1}{24}\tau^4+ ...\right). \end{align}
 Omitting further
calculations, we present the final result for the scalar
Lorentz-Dirac-DeWitt-Brehme force: $$ \label{6D14} m_0
\ddot{z}^{\mu}=m^2 q^2\left(f^{\mu}_{\rm flat}+f^{\mu}_{\rm curv
\; div}+f^{\mu}_{\rm curv  \; fin}+f_{\rm tail}^{\mu}\right),
$$
where the ``flat-space'' part is given by
 \begin{align} \label{6D15} f^{\mu}_{\rm
flat}=& \frac{1}{6\varepsilon^3}
\ddot{z}^{\mu}+\frac{1}{\varepsilon}\left(-\frac{1}{16}\ddot{z}^2
\ddot{z}^{\mu}+\Pi^{\mu \nu}\frac{1}{24}z^{(4)}_{\nu}\right)-\nnn
&
 -
\frac{2}{9}(\stackrel{...}{z}\ddot{z})\ddot{z}^{\mu}-\Pi^{\mu
\nu}\left[\frac{1}{9}\ddot{z}^2\stackrel{...}{z}_{\nu}+\frac{1}{45}z^{(5)}_{\nu}\right].
\end{align}It has the same form as in the Minkowski space, but the dots
denote covariant derivatives along the world-line. It includes two
divergent terms proportional to $\varepsilon^{-3}$ and
$\varepsilon^{-1}$, and it is the only part of the total force
which survives in the flat space limit.

The divergent part induced by the curvature contains only
$\varepsilon^{-1}$ term and is given by \begin{align} \label{6D16}
f^{\mu}_{\rm curv \; div}=& -\frac{1}{72 \varepsilon}\Pi^{\mu \nu}
\biggl(2R_{\nu\alpha}\ddot{z}^{\alpha}+2
R_{\nu\alpha;\,\beta}\dot{z}^{\alpha}\dot{z}^{\beta}-\nnn & -
R_{\alpha\beta;\,\nu}\dot{z}^{\alpha}\dot{z}^{\beta}+
R_{\alpha\beta}\ddot{z}_{\nu}\dot{z}^{\alpha}\dot{z}^{\beta}+
R_{;\,\nu}+R \ddot{z}_{\nu}\biggr). \end{align} The leading
divergence $\varepsilon^{-3}$ is absorbed by the renormalization
of mass,
\begin{equation}m=m_0-\frac{m_0^2 q^2}{6\varepsilon^3},\end{equation}
it is not affected by the curvature. To eliminate the second
divergence $\varepsilon^{-1}$ one has to introduce the following
counterterm into the action: \begin{align}\label{6D17}
 S_c^{\rm sc}  =  \frac{\kappa^0}{72}  \int &\!
\left( \frac{3}{2}(\tilde{D}^2 z)^2 +\right. \nnn & \!+
\left(R-R^{\alpha\beta}Dz^{\alpha}Dz^{\beta}\right)\!\biggr)
\sqrt{-\dot{z}^2}d\tau,
\end{align}where $\kappa_0$ is a new bare coupling constant, and
$\tilde{D}=1/\sqrt{-\dot{z}^2}\cdot D/d\tau$ is the
reparameterization-invariant covariant derivative along the
world-line. The first term in $S_c$ is a covariantization of the
one obtained in the flat space \cite{Ko99}, while the second
Ricci-dependent term is new. Note that it vanishes in the region
of space-time free of matter sources. The bare coupling constant
$\kappa^0$ entering the counterterm is renormalized to
\begin{equation} \label{6D18}
\kappa=\kappa^0-\frac{m_0^2 q^2}{\varepsilon}.
\end{equation}
Non-vanishing $\kappa$ leads to rigid particle dynamics
investigated in a number of papers (see \cite{Ko07} for the
references).

The local part of the finite force induced by the curvature depends
both on the Ricci and the Riemann tensor, it does not vanishes
therefore in the vacuum region of the space-time: \begin{widetext}
\begin{align} \label{6D20} & f^{\mu}_{\rm curv\;fin}=\Pi^{\mu
\nu}\left[\left(-\frac{1}{45}\,R^{\rho \;\; \tau}_{\;\;\nu
\;\;\alpha}R_{\rho \beta\tau \gamma}- \frac{1}{36}\,R_{\nu \alpha
}R_{\beta \gamma}-  \frac{1}{180}\,R_{\alpha}^{\rho }R_{\rho \beta
\nu \gamma}-\frac{1}{20}\,R_{\nu \alpha ;\, \beta
\gamma}+\frac{1}{30} \,R_{\alpha \beta; \, \nu
\gamma}\right)\dot{z}^{\alpha }\dot{z}^{\beta}\dot{z}^{\gamma}
\right.\nnn & +\left(\frac {1}{360}R_{\nu}^{\;\; \rho \tau
\lambda}R_{\alpha \rho \tau \lambda}+\frac{1}{180}\,R^{\lambda\rho
}R_{\lambda\nu \rho \alpha }- \frac {1}{60}\,R_{;\, \nu \alpha}-
-\frac {1}{90}\,R_{\nu}^{\lambda}R_{\alpha \lambda}+\frac
{1}{72}\,R_{\nu \alpha}R + \frac {1}{120}\,\Box R_{\nu
\alpha}\right)\dot{z}^{\alpha}- \nnn & - \frac{1}{6}\,R_{\alpha
\beta}
\ddot{z}^{\alpha}\dot{z}^{\beta}\ddot{z}_{\nu}-\frac{1}{18}\,R_{\nu
\alpha} \stackrel{...}{z}^{\alpha}+\frac{1}{36}\,R
\stackrel{...}{z}_{\nu} -  \frac{1}{9}\,R_{\alpha
\beta}\dot{z}^{\alpha
}\dot{z}^{\beta}\stackrel{...}{z}_{\nu}-\frac{1}{12}\,R_{ \alpha
\beta,\gamma}\dot{z}^{\alpha
}\dot{z}^{\beta}\dot{z}^{\gamma}\ddot{z}_{\nu} +\frac{1}{18}\,R_{\nu
\alpha }\ddot{z}^2\dot{z}^{\alpha }-  \nnn & -\frac{1}{12}\,R_{\nu
\alpha ; \,\beta}\ddot{z}^{\alpha}\dot{z}^{\beta}
\left.-\frac{1}{12}\,R_{\nu \alpha ; \,\beta}\dot{z}^{\alpha
}\ddot{z}^{\beta}+ \frac{1}{12}\,R_{\alpha \beta;
\,\nu}\dot{z}^{\alpha }\ddot{z}^{\beta}\right]. \end{align}
\end{widetext}In the case of the geodesic motion the last four lines in
this expression vanish. The non-vanishing part is quadratic in
curvature.

The non-local tail term is given by
\begin{equation} \label{6D21}
 f_{\rm
tail}^{\mu}=-\frac{1}{3}\Biggl(\Pi^{\mu\nu}\int\limits_{-\infty}^{\tau}v_{
\nu} d\tau'+\ddot{z}^{\mu}\int\limits_{-\infty}^{\tau}v
d\tau'\Biggr).
\end{equation}

\subsection{Vector field}\label{6Dcurveem}
In six dimensions the Hadamard function is
\begin{equation} \label{6D1em}
\!\!\!\!
G^{\nu\alpha}_H=\frac{1}{(2\pi)^3}\left[\frac{\Delta^{1/2}\bar{g}^{\nu\alpha}}{\sigma^2}+
\frac{u^{\nu\alpha}}{\sigma}+v^{\nu\alpha}\ln
\sigma+w^{\nu\alpha}\right],
\end{equation}
while the retarded Green's function reads: \begin{align}
\label{6D3em} G&_{\rm  ret}=
-\frac{1}{8\pi^2}\theta(x',\Sigma(x))\times \nnn & \times
\left[\Delta^{1/2}\bar{g}^{\nu\alpha}\delta'(\sigma)-u^{\nu\alpha}\delta(\sigma)+
v^{\nu\alpha}\theta(-\sigma)\right].\! \end{align} The retarded
potential will be \begin{align} \label{6D3aaaem}
 A^{\nu}_{\rm ret}(x)=\frac{e}{3} \! \int\limits_{-\infty}^{\tau_{\rm
 ret}(x)}\!
& \left[ \Delta^{1/2}\bar{g}^{\nu}_{\;\,\alpha}\delta'(\sigma)-
u^{\nu}_{\;\,\alpha}\delta(\sigma)\right. \nnn & \;+ \left.
v^{\nu}_{\;\,\alpha}\theta(-\sigma)
\vphantom{\Delta^{1/2}}\right]\dot{z}^{\alpha} d\tau'.
\end{align}
Omitting the details of the calculation we present the final
result in the form
\begin{equation} \label{6D14dem}
 m_0 \ddot{z}^{\mu}=e^2\left( f^{\mu}_{\rm
flat}+f^{\mu}_{\rm curv \, div}+f^{\mu}_{\rm curv  \,
fin}+f^{\mu}_{\rm tail}\right),
\end{equation}
where again the ``flat'' part is the covariantization of the
flat-space expression \begin{align} \!f^{\mu}_{\rm flat} \! =&
-\frac{1}{6\varepsilon^3}
\ddot{z}_{\nu}+\frac{1}{\varepsilon}\Pi^{\mu\nu}\!\left(-\frac{3}{16}\ddot{z}^2
\ddot{z}_{\nu}+\frac{1}{8}z^{(4)}_{\nu}\right) \nnn & +
\frac{2}{3}(\stackrel{...}{z}\ddot{z})\ddot{z}^{\mu}-\Pi^{\mu\nu}\!\left(\frac{4}{45}z^{(5)}_{\nu}+
\frac{2}{9}\ddot{z}^2 \stackrel{...}{z}^{\mu}\!\right).\!
\end{align}The divergent term is \begin{align} \label{6D16em}
& f^{\mu}_{\rm curv \; div}=-\frac{5}{72 \varepsilon}\Pi^{\mu \nu}
\biggl( 2R_{\nu\alpha}\ddot{z}^{\alpha}+2
R_{\nu\alpha;\,\beta}\dot{z}^{\alpha}\dot{z}^{\beta}-\nnn &
 -
R_{\alpha\beta;\,\nu}\dot{z}^{\alpha}\dot{z}^{\beta}+
R_{\alpha\beta}\ddot{z}_{\nu}\dot{z}^{\alpha}\dot{z}^{\beta}+
\frac{1}{5}(R_{;\,\nu}+R \ddot{z}_{\nu})\biggr)\!. \end{align} The
leading divergence is absorbed by the mass renormalization, while
the non-leading one is eliminated introducing the following
counterterm: \begin{align} \label{6D17emmm}
 S_c^{\rm em}  =  \frac{\kappa^0}{72} \! \int\!\!&
\left( \frac{9}{2}(\tilde{D}^2 z)^2- \right. \nnn & -
(R+5R^{\alpha\beta}Dz^{\alpha}Dz^{\beta})\!\biggr)\!
\sqrt{-\dot{z}^2}\, d\tau.\! \end{align}The finite part of the
self-force $f^{\mu}_{\rm curv  \, fin}$ can be cast into the sum
of  the following parts: a) the quadratic in curvature terms:
\begin{align} \label{emquad} f_{\rm quad}^{\mu}
=& \Pi^{\mu\nu}\! \left(\frac{1}{36}
 R_{\nu\sigma}R_{\tau\rho}{\dot{z}}^{\sigma}{\dot{z}}^{\tau}{\dot{z}}^{\rho}+
 \frac{7}{864}R\,R_{\nu\sigma}{\dot{z}}^{\sigma}
 \right. +\nnn & \quad \;\;+\left.
 \frac{1}{1080}R_{\nu\rho\sigma\tau}R^{\alpha\rho\sigma\tau}{\dot{z}}_{\alpha}+
 \frac{29}{4320}R_{\nu\sigma}R^{\sigma\tau}{\dot{z}}_{\tau} \right. \nnn &\quad\;\; +\left.
 \frac{1}{540}R_{\nu\rho\sigma\tau}R^{\rho\tau}{\dot{z}}^{\sigma}\right);
\end{align} b) the Ricci-terms:
\begin{align} \label{emRic}
  f_{\rm Ric}^{\mu}&  =\Pi^{\mu\nu}\left(\frac{4}{27}R_{\nu\sigma}\stackrel{...}{z}^{\sigma}-
 \frac{1}{27}\ddot{z}^2 R_{\nu\sigma}\dot{z}^{\sigma}
 \right. \nnn & \left.+\frac{2}{27}R_{\sigma\tau}\dot{z}^{\sigma}{\dot{z}}^{\tau}
 \stackrel{...}{z}^{\nu}+\frac{2}{9}R_{\nu\sigma ;
 \tau}\ddot{z}^{\sigma}{\dot{z}}^{\tau}+\frac{1}{60}\Box R_{\nu\sigma}{\dot{z}}^{\sigma}\right.
 \nnn &
 \left.+\frac{1}{18} R_{\nu\sigma ;
 \tau}\ddot{z}^{\tau}{\dot{z}}^{\sigma}-\frac{1}{18} R_{\sigma \tau  ;\,
 \nu}{\dot{z}}^{\sigma}\ddot{z}^{\tau}\right. \nnn & \left.+\frac{1}{12}R_{\nu\sigma;\,\rho\tau}{\dot{z}}^{\sigma}{\dot{z}}^{\tau}{\dot{z}}^{\rho}-
 \frac{1}{36}R_{\rho\sigma;\,\nu\tau}{\dot{z}}^{\sigma}{\dot{z}}^{\tau}{\dot{z}}^{\rho}
 \right);
\end{align}c) the Ricci-scalar terms:
$$
  f_{\rm Rsc}^{\mu}\!
 =\Pi^{\mu\nu}\left(-\frac{1}{54}R \stackrel{...}{z}_{\nu}-\frac{1}{36}R_{;\sigma}{\dot{z}}^{\sigma}\ddot{z}_{\nu}-
 \frac{1}{72}R_{;\,\nu\sigma}{\dot{z}}^{\sigma}
 \right)\!;
$$
d) the Riemann term
\begin{equation} \label{emRiem}
 f_{\rm Riem}^{\mu}
 =-\frac{1}{9}R^{\mu}_{\;\;\sigma\nu\tau}{\dot{z}}^{\nu}{\dot{z}}^{\sigma}\stackrel{...}{z}^{\tau}.
\end{equation}
The appearance of the Riemann tensor in the finite local force in
a new feature in six dimensions, which is absent in the scalar
case. This term is non-zero in the vacuum space-time, but it still
vanishes for geodesic motion.

Finally, we observe that the tail term has the same form as in
four dimensions.

\section{Conclusions} \label{concl}
In this paper we presented a purely local calculation of the
radiation reaction for a point particle in curved space of an even
dimension $D\geqslant 4$. The possibility of such a calculation
means that scattering of radiation on curvature in the vicinity of
a charge is irrelevant for the reaction force. Clearly the
scattering can substantially modify global properties of
radiation. In particular, emitted waves can be reflected back and
act on a particle once again, but within our approach this should
be rather interpreted as an action of the external field, than a
modification of the reaction force. Note, however, that such a
description may not be natural if the problem is treated globally
in terms of the field modes defined on the full manifold. In
particular, for space-times with reflection of massless fields
from infinity (like anti-de Sitter) the action of modes (subject
to reflection conditions) on a particle will include the effect of
scattering, thus leading to a global definition of the reaction
force. In terms of the above analysis this means that there are
several geodesics connecting two points of space-time. Our
approach is valid only under the assumption that this is not the
case. It excludes, therefore, many physically interesting
situations, which require a special global analysis. For an
example of the global definition of the reaction force see
\cite{kerry} where the radiation reaction problem was considered
for particles moving in the Kerr space-time, (for the case of a
cosmic string see \cite{Ga90,Khus05}).

Another point worth to be discussed is that of elimination of
divergences. Classical renormalizability of the equations for
point particles interacting with massless scalar and vector fields
in flat space time is well-known for $D=4$. In higher dimensional
Minkowski space it also holds modulo introduction of the higher
derivative counterterms into the action \cite{Ko07,Ko99}. Here we
have demonstrated that this remains valid in the general curved
space-time of an arbitrary dimension for a minimally coupled
linear scalar and vector fields. The curved space counterterms are
not simply the covariant generalizations of the flat space ones,
but generically include curvature-dependent terms. An explicit
example was given for the $D=6$ case. One observes that the
renormalized higher-dimensional reaction force contains not only
higher derivatives of the particle velocity, but also higher
derivatives of the Ricci tensor, as well as the Riemann tensor, a
feature which is not present in four dimensions.

\end{document}